# AUTOMATIC EXTRACTION OF REQUIREMENTS EXPRESSED IN INDUSTRIAL STANDARDS : A WAY TOWARDS MACHINE READABLE STANDARDS ?

**Hélène de Ribaupierre**
Cardiff University, UK
**Anne-Françoise Cutting-Decelle**
Université de Genève/CUI, CH
**Nathalie Baumier, Serge Blumental**
RTE, F

**Abstract** : The project, under industrial funding, presented in this publication aims at the semantic analysis of a normative document describing requirements applicable to electrical appliances. The objective of the project is to build a semantic approach to extract and automatically process information related to the requirements contained in the standard. To this end, the project has been divided into three parts, covering the analysis of the requirements document, the extraction of relevant information and creation of the ontology and the comparison with other approaches.

The first part of our work deals with the analysis of the requirements document under study. The study focuses on the specificity of the sentence structure, the use of particular words and vocabulary related to the representation of the requirements. The aim is to propose a representation facilitating the extraction of information, used in the second part of the study.

In the second part, the extraction of relevant information is conducted in two ways: manual (the ontology being built by hand), semi-automatic (using semantic annotation software and natural language processing techniques). Whatever the method used, the aim of this extraction is to create the concept dictionary, then the ontology, enriched as the document is scanned and understood by the system. Once the relevant terms have been identified, the work focuses on identifying and representing the requirements, separating the textual writing from the information given in the tables. The automatic processing of requirements involves the extraction of sentences containing terms identified as relevant to a requirement. The identified requirement is then indexed and stored in a representation that can be used for query processing.

## Introduction

Requirements analysis is an area that has been studied for a long time and has resulted in a large number of research publications, in particular as applications of NLP (Natural Language Processing) [1]. The developments made generally concern the way of processing requirements, when they are expressed in textual form -- natural language plays an important role in identifying and extracting the information contained therein [2], [3], [4]. The fields of application of requirements analysis are very varied, whether in software engineering, in the legal field or in the industrial sector [5].

In the field of technical standardization, requirements play an important role, as a large number of normative documents simply translate technical requirements for systems or products into standards, conversely, standards can often be directly integrated into technical requirements. However, there seems to be little research in the area of requirements analysis and processing in normative documents in the industrial domain.

Several actions are currently starting either at the EU level (CEN / CENELEC) or at the international level (IEC, ISO), such as : Smart Standards, Standards of the Future, all of them aimed at developing the concept of Machine Readable Standards (MRS) [6], [7]. The study presented here, object of the project, can be seen as a contribution towards the MRS approach.

The purpose of the RTE-DIGIREQ project described in this paper is the semantic analysis of a normative document, in this case the IEC 60376-2018 standard, "Specification of technical grade sulphur hexafluoride (SF6) and complementary gases to be used in its mixtures for use in electrical equipment " [8], from the point of view of the study of the requirements defined in the standard. The objective is to build a semantic approach for the extraction and automatic processing of the information related to the requirements in the normative document. This semantic analysis of a normative document is based on natural language techniques.

The work subject of the project was initially done on the two versions of the standard : the English version and the French version, however only the English version is presented in this paper.

The structure of the paper is the following : in the first part, we present the concept of requirement, particularly in its application to normative documents. The second part will provide an overview of the ISO 60376 standard, subject of the study. We will then propose a first study for an automatic extraction of requirements, through the description of the overall approach and the method followed. The following section will describe the process of extraction and processing of sentences using a text engineering approach, thus leading towards an automation of requirements formalisation. The paper will end with the proposal of some issues and perspectives for the future.

**1 The concept of requirement – application to normative documents**

**1.1 Concept of requirement**

Among the different definitions of a requirement, one of the most commonly accepted is : « a requirement is the expression of a condition or functionality that a system or software must meet ». (SPECIEF) [9]

The definitions used by the ISO, IEEE, CMMI standards are:

1) Condition or capability that a user needs to solve a problem or achieve a goal ;
2) Condition or capability that a product or component of a product must possess to fulfil a contract, conform to a standard, specification or other formally imposed document ;
3) Documented representation of a condition or capability as in (1) or (2).

**1.2 Types and representations of requirements**

There are different types of requirements, corresponding to the different approaches of a system and the ways of expressing requirements. The Business Analysis Body of Knowledge (BABOK [10]) defines the following requirements types : Architectural requirements, Business requirements, User (stakeholder) requirements, Functional (solution) requirements, Quality-of-service (non-functional) requirements, Implementation (transition) requirements, Regulatory requirements. Requirements can be represented as diagrams, tree structures, or textual documents (specifications – including or not schemas, diagrams).

For the project presented in this paper, since the focus of the study was the extraction of requirements expressed in an International Standard, we are mainly interested in requirements represented in textual form, either in plain text or in the form of tables of data or numerical values.

**1.3 Requirements and normative documents**

Requirements and their representation provide an important contribution to standardization activities, particularly for the standards applicable to the technical and industrial sectors.

Requirements are described in the ISO/IEC Directives Part 2 on writing normative documents (ISO/IEC Directives, Part 2) [11] : for the Directives, a requirement is an « *expression, in the content of a document, that conveys objectively verifiable criteria to be fulfilled and from which no deviation is permitted if conformance with the document is to be claimed* ».

The way of expressing requirements is of the highest importance for the standards, particularly through the use of verbs, and verbal forms : requirements, as well as recommendations, possibilities and capabilities, external constraints are expressed using verbal forms (e.g. use of "shall", "shall not") specified in the Directives document. According to the Directives, « *it is essential to follow rules for the use of verbal forms so that a clear distinction can be made between requirements, recommendations, permissions, possibilities and capabilities. To avoid risk of misinterpretation, verbal forms that are not defined in* (the corresponding tables) *shall not be used for the expression of provisions* ».

For the ISO/IEC/IEEE 29148:2018 standard [12], a « *requirement is a statement that translates or expresses a need and its associated constraints and conditions. A requirement can be written in the form of a natural language or some other form of language. If expressed in the form of a natural language, the statement should include a subject and a verb, together with other elements necessary to adequately express the information content of the requirement. A requirement shall state the subject of the requirement (e.g., the system, the software, etc.), what shall be done (e.g., operate at a power level, provide a field for) or a constraint on the system. Fig. 1 below shows an example syntax for requirements.* »

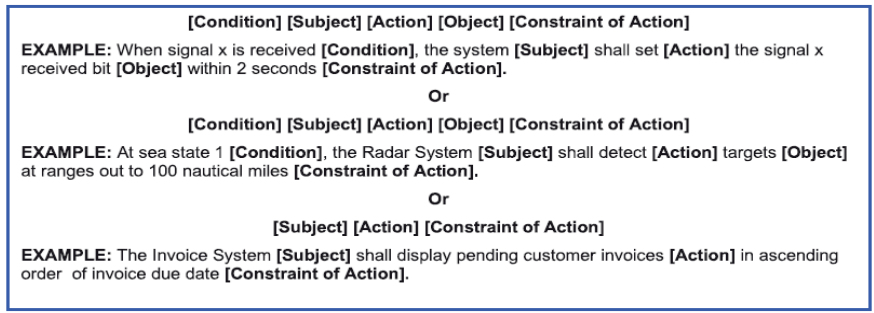

Fig. 1 : Examples of functional requirements syntax (ISO/IEC/IEEE 29148:2018)

For the project and the analysis of the requirements, we focused on the IEC 30676 standard, "Specification of technical grade sulfur hexafluoride (SF6) and complementary gases to be used in its mixtures for use in electrical equipment".

**2 The IEC 60376 standard : scope and content in terms of requirements**

**2.1 Description of the IEC 60376 standard**
The scope of the standard, developed by the IEC TC 10 committee (Fluids for electrotechnical applications) « *defines the quality for technical grade sulphur hexafluoride (SF6) and complementary gases such as nitrogen (N2) and carbon tetra-fluoride (CF4), for use in electrical equipment. Detection techniques, covering both laboratory and in-situ portable instrumentation, applicable to the analysis of SF6, N2 and CF4 gases prior to the introduction of these gases into the electrical equipment are also described in this document* ». (IEC FDIS 60376-2018)
Sulphur hexafluoride (SF6) is an artificial gas widely used in high voltage electrical equipment. It is colourless, odourless, non-combustible and chemically very stable. It does not react with other substances at room temperature. Its high stability is based on the perfect symmetrical arrangement of its six fluorine atoms around its central sulphur atom ; it is precisely this stability that makes this gas so useful in industry. SF6 is an excellent electrical insulator and can effectively extinguish an electric arc. This has made it very popular and today thousands of medium and high voltage electrical equipments worldwide use it. The specifications of the SF6 gas to be used in electrical equipments are described through a set of requirements which are the subject of the standard.

The study presented in this document focuses on the way to automate the processing of requirements. However if this study focuses on the particular set of requirements described in the standard subject of the project, the method proposed could be used for different types of requirements.

**2.2 Examples of requirements mentioned in the standard**
In this standard, there are mainly two expressions of the requirements : one is written in natural language (English for the standard analysed) and the other is a mix of text in a free environment and text in tables (with numerical data and expressions). Let us take an example of each :
- Ex 1 : Requirement for responsibility (text) : *It is the responsibility of the supplier to guarantee that the delivered gas or gas mixture is non-toxic, in accordance with international and local regulations.*
- Ex 2 : Requirement for storage and transportation (text) : *Information concerning gas storage and transportation is provided in IEC 62271-4. Specific labelling of containers shall be implemented in accordance with the mode of transport and the local and international regulations.*
- Ex 3 : Requirements for complementary gases to be used in SF6 mixtures (described in 2 tables) : *$SF_6$ mixtures are used in electrical equipment mainly for cold ambient temperature applications, typically under -40 °C. Other applications at normal ambient temperature include gas insulated transmission lines (GIL) and gas insulated transformers (GIT). $SF_6$ is mixed with a complementary gas, typically $N_2$ or $CF_4$, in the percentage as specified by the original equipment manufacturer in the operating instruction manual, typically from 10 % to 75 % $SF_6$ volume. The maximum permitted concentrations of other substances present in $N_2$ are given in Table 2 and in Table 3 for $CF_4$.*

The analysis presented here only focuses on the Table 3 of the requirement related to the complementary gases to be used in SF6 mixtures.

**Table 1 : [Table 3 – Requirements for CF4 to be used in SF6 mixtures (IEC 60376:2018)]**

| Substance | Concentration |
|---|---|
| $CF_4$ | > 99,7 % volume |
| $O_2$ | < 500 µl/l (i.e. 500 ppmv) |
| $N_2$ | < 1 500 µl/l (i.e. 1 500 ppmv) |
| $H_2O$ | < 200 µl/l (i.e. 200 ppmv) |
| Mineral oil | < 10 mg/kg (i.e. 10 ppmw) |
| Total acidity | < 7 µl/l (i.e. 7 ppmv) |

**Key** ppmv = parts per million by volume ppmw = parts per million by weight

\*\* Notes, comments on the requirements mentioned in the IEC 60376 standard :
1- Requirement about responsibility (section 4 of the standard, General requirements) : the concepts of "*responsibility*" (in the context of the requirements) and "*guarantee*" do not appear in the Directives : although they correspond to a legal approach often encountered in the formulation of requirements.
2- Requirement on the environmental impact (« *SF6, CF4 and SF6 mixtures with N2 and/or CF4 have a certain environmental impact. Due to this impact, SF6, CF4 and their mixture gas shall be handled carefully to prevent deliberate release of SF6 and CF4 gas into the atmosphere.* ») : there is an confusion in the terms used : what does « *certain* » mean -- same for « *carefully* » and « *deliberate* » ?   -- these terms will require a special treatment before they can be considered for recognition as important terms in a requirement.
3- Storage and transportation : what does « *in accordance with* … » mean in this context ?  -- same for « *specific* »

In the following sections, we will present the overall approach enabling an automatic extraction of requirements, then the application of the methodology to the examples of requirements listed above.

**3. Towards an automatic extraction of requirements: overall approach and method**

In this section, we present the approach developed in order to automate the processing of requirements, then some results in terms of extraction (section 4). It should be noted that this is only a first approach which needs to be followed by further work.

The approach we choose here is based on the use of domain ontologies or glossaries to detect the extraction of the relevant concepts in a sentence. This approach allows us to have a very precise control on the selection of the concepts that we are extracting. However, this approach has also its "bad side", as if the knowledge base doesn't exist, we need to create one. In a second step, we need to identify the sentences that are containing the requirements and the concepts identified in the previous step. In addition of detecting the concept contained in the domain of the requirement, we will need to detect the pattern of the requirement, that will help us to automatically determine a sentence that contains a requirement from another sentence. In this work, this detection was made manually.  In a third step, after detecting the sentences containing a requirement, we will need to transform the sentences in a formal language. In this work, we chose the RSML  (Requirements Specific Modeling Language) language [13], [14]. The last step is to store those sentences in a database, allowing the user to query the results and visualize them. We will detail these different steps in the following sections.

**3.1  Identification of the relevant concepts and structuring (glossary, then ontology)**
**3.1.1 Glossary of the relevant terms** : we consider as « relevant » the terms appearing in the requirements as formulated in the standard, either in the text part, or in the property tables. The degree of relevance we are interested in here is therefore quite small, since it only concerns the meaning of the term in the context of the standard.

Note: we may find in this list terms that are used with a different meaning from the one they have in the current vocabulary: we will list them, and we will then have to ask ourselves the question of the semantic treatment to be applied to them.

For the moment, the ontologies developed are not very axiomatised, only a few properties are mentioned insofar as they appear in the requirements. A later development will take into account these differences in the meanings of the terms in relation to the current vocabulary, in order to avoid misinterpretation. Those terms are presented in a bilingual glossary (since we worked on the two versions, english and french of the standard). It may be noted, however, that some terms do not exist in both languages, for other, the translation proposed is not similar to the translation usually proposed in the dictionaries. The glossary is available as an Excel file.
If, at first, a glossary may seem sufficient to extract, once identified, the information contained in the requirements - after having first identified the sentences/tables likely to be treated as requirements -, the structuring in the form of an ontology is necessary to analyse the requirements with language processing tools that do semantic annotation. This is the way to identify, group and align terms in the text that are subject to a similar semantic definition. Besides, it is essential in the case of the development of an automatic requirements processing tool.

**3.1.2 References to existing ontologies/dictionaries/thesaurus** : as far as possible, we have made investigations on the existing information sources of the domain, such as, for our example, chemical ontologies [15]. This analysis will be very important to allow interoperability between the requirement extraction approach of the standard under study and other normative documents in the domain or in neighboring domains, especially in the context of machine readable standards (MRS).

**3.1.3 Structuring of the terms (ontology**) : ontology of the concepts related to chemistry used in ISO 60376. Two versions of the ontology have been developed, one in French, the other in English. Fig. 2 below shows an excerpt of the english version of the ontology and Fig. 3 shows the structuring of the concept of properties in the ontology (Ontograf schema), as represented using the Protégé 5.5 software [16].

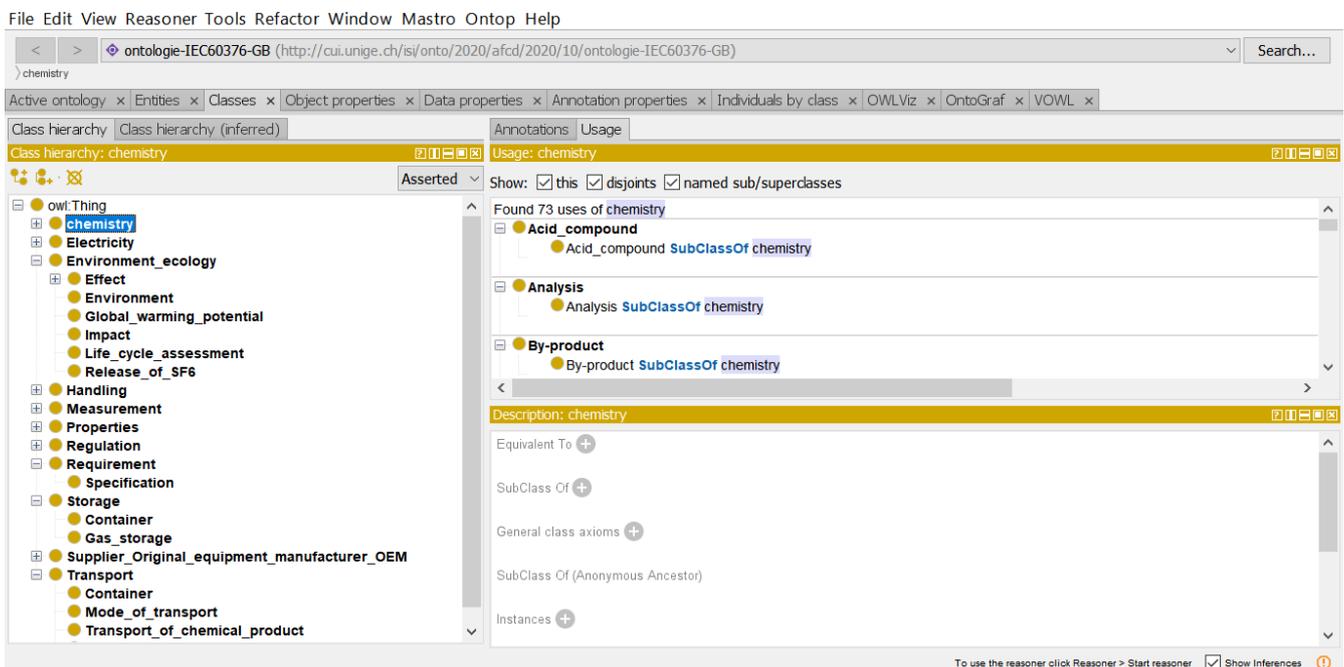

Fig. 2 : English version of the ontology of IEC 60376 (excerpt)

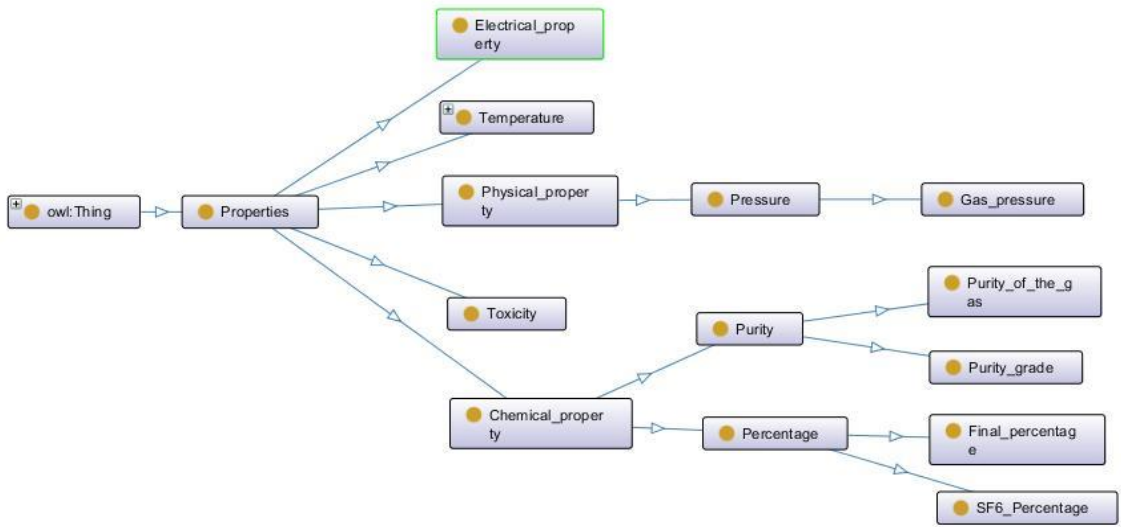

Fig. 3 : Structure of the concept of « properties » (Ontograf representation) - Excerpt

The following sections will show how the use of the ontology will be used to help the extraction of the requirements from the standard.

**3.2 Extraction of sentences**
In this section, we focus on the extraction of sentences. Sentence extraction (in documents and/or corpora) in order to exploit significant information has been the subject of a large body of research, including [17] [18], showing the different methods that can be used. In this work, we did choose GATE[1] [19] as this is a well-known open-source framework. GATE provides an architecture and a framework environment to develop and deploys natural language software components. It offers a rich graphical user interface, and provides an easy access to language, processing and visual resources that help scientists and developers to produce a natural language processing application.
JAPE (Java Annotation Pattern Engine) is a finite state transducer, using regular expressions for handling pattern-matching rules. Such expressions are at the core of every rule-based IE system aimed at recognizing textual snippets that conform to patterns while the rules enable a cascading mechanism of matching conditions that is usually referred as the IE pipeline. JAPE grammars are constituted from two parts; the LHS (Left Hand Side) which handles the regular expressions and the RHS (Right Hand Side) which manipulates the results of the matching conditions and defines the semantic annotation outcome.
The GATE family of tools has grown over the years to include a desktop client for developers, a workflow-based web application, a Java library, an architecture and a process. GATE as an architecture suggests that the elements of software systems that process natural language can usefully be broken down into various types of components, known as resources. In addition, the platform offers easy way of loading ontologies and annotating text using these ontologies.

**3.3 Modelling approach**
We focus here on the formal approaches that offer the possibility of moving seamlessly from the expression of requirements to the implementation of the code that satisfies them, possibly using a pivot language to facilitate the transition from the requirements language to the computer code [20]. Requirements modelling (especially in the field of software development) is a well-studied area, which has given rise to a great deal of research work that has led to the development of several approaches. Some of these approaches have subsequently been developed commercially, others have remained open-source.
For the project, we considered some of them, such as RSML, ReqIF [21], EARS [22]. We decided to focus on RSML since it seemed to us the most suited to the needs of the project.

RSML[2] (Requirements Specific Modeling Language) [13], [14] is a modeling language specific to requirements modeling that allows requirements to be expressed in a grammar close to the natural language. The editor has been defined with the help of the Gemoc[3] studio. The use of a model-based approach allows to combine it with other languages such as SysML or Eiffel. For RSML, the author has chosen to give a formal definition in Eiffel. RSML proposes a seamless approach, from the elicitation of requirements to the documentation of an implementation, by proposing a concrete syntax closer to natural language, while using the semantics of Eiffel.

---
[1] https://gate.ac.uk
[2] https://gitlab.com/fgalinier/RSML

- Application to the Ex 1 (Requirement for responsibility (text)) : *It is the responsibility of the supplier to guarantee that the delivered gas or gas mixture is non-toxic, in accordance with international and local regulations.* The RSML code for the requirement is :

```
Environment:
- Toxicity of the gas is in 0 and 100 .
- Toxicity of the gas mixture is in 0 and 100 .
- Responsibility of the supplier is in 0 and 1 .
SUPPLIER:
[1]"The supplier provides the gas ."
[2]"The supplier provides the gas mixture ."
RESPONSIBILITY OF THE SUPPLIER:
[3] When the toxicity of the gas is equal to 0 then immediately the responsibility of the supplier shall be equal to 0 .
INTERNATIONAL REGULATION:
[4] According to international regulation the toxicity of the gas shall be equal to 0 [Percentage] .
[5] According to international regulation the toxicity of the gas mixture shall be equal to 0 [Percentage] .
LOCAL REGULATION:
[6] According to local regulation the toxicity of the gas shall be equal to 0 [Percentage] .
[7] According to local regulation the toxicity of the gas mixture shall be equal to 0 [Percentage] .
```

A certain number of concepts relevant to the scope of the standard appear in the requirements : to make the requirements extraction process automatic, it is necessary to identify those concepts, than to structure them : this identification (glossary), then the structuring (ontology), are presented in the following paragraph.

### 3.4 Method used for the identification and representation of the requirements

**- Identification** : the identification of requirements is done here first by hand, whether for the text or for the tables, in order to be able to show, when there is one, the logic of the drafting and then facilitate the transformation allowing an automatic extraction/reading of the requirements. The difficulty of the problem lies in the way of isolating a requirement : generally speaking, we consider that a sentence (or a table) can be assimilated to a requirement as soon as it contains one of the words that we have identified as meaningful concepts (ontology) associated with verbs such as "SHALL", "MUST".

This should be supported by the rules for drafting standards – however it is not always the case, the problem is not simple, with sometime ambiguïties in the documents, as we will see at the end of this paper.

It is these criteria that we will enter in the form of rules (JAPE) in the semantic annotation software.

**\* Comments :**
   - a sentence can give rise to several requirements, in particular to deal with the coordinating conjunctions that appear in the text: "AND", "OR", in order to take into account the complexity of certain formulations and to avoid the ambiguities that can occur in the formulation of natural language.
   - The formulation of the requirements represented below can be seen as a form of pseudo-code. This representation has the advantage of simplifying the initial syntax of the sentence and of proposing only one proposition per line.

We will see later that, depending on the language used for modelling the requirements, we shall have to rework the sequence of words in the sentence.

**- Representation :** we present here the breakdown of the sentence describing a requirement : tokens representing concepts and terms specific to the vocabulary of the requirement (Distinguish_features column) :

\* Distinguish feature: represents something that is unique and shows that it is a requirement, example: SHALL
\* Concepts 1, 2, 3, 4: terms represented in the ontology.

When a sentence is too complex, it is broken down into simpler elements in the requirement model.

We limit here to the representation of requirements expressed in text form. Requirements expressed in tabular form must be treated separately.

---

[3] http://gemoc.org/studio.html

- Ex 1 : Requirement for responsibility (text) : *It is the responsibility of the supplier to guarantee that the delivered gas or gas mixture is non-toxic, in accordance with international and local regulations.*

| Sentences | Distinguish_features | Concept1 | Concept2 | Concept3 | Concept4 |
|---|---|---|---|---|---|
| It is the responsibility of the supplier to guarantee that the delivered gas or gas mixture is non-toxic, in accordance with international and local regulations. | GUARANTEE<br><br>in conformance with :<br>(international regulation<br>    and   local regulation) | Supplier | Gas<br>or<br>gas mixture | toxicity = 0 | |

- Ex 2 : Requirement for storage and transportation (text) : *Information concerning gas storage and transportation is provided in IEC 62271-4. Specific labelling of containers shall be implemented in accordance with the mode of transport and the local and international regulations.*

For each of the sentences describing the requirement, the relevant terms are extracted, then represented in the table below :

| Sentences | Distinguish_features | Concept1 | Concept2 | Concept3 | Concept4 |
|---|---|---|---|---|---|
| Information concerning gas storage and transportation is provided in IEC 62271-4.<br>Specific labelling of containers shall be implemented in accordance with the mode of transport and the local and international regulations. | SHALL<br><br>local regulation<br>international regulation | container labelling | storage | mode of transport | transport |

- Ex 3 : Requirements for complementary gases to be used in SF6 mixtures (2 tables) : *$SF_6$ mixtures are used in electrical equipment mainly for cold ambient temperature applications, typically under -40 °C. Other applications at normal ambient temperature include gas insulated transmission lines (GIL) and gas insulated transformers (GIT). $SF_6$ is mixed with a complementary gas, typically $N_2$ or $CF_4$, in the percentage as specified by the original equipment manufacturer in the operating instruction manual, typically from 10 % to 75 % $SF_6$ volume. The maximum permitted concentrations of other substances present in $N_2$ are given in Table 2 and in Table 3 for $CF_4$. – See Table 1.*

The expression of the requirement as expressed from the table is represented below :

**EXPRESSION OF THE REQUIREMENT (table) :**
- **substance = N2 :**
**Concentration GREATER THAN 99,7 % by volume**
- **substance = H2O :**
**Concentration SMALLER THAN 200 µl/l (= 200 ppmv)**
- **substance = O2 :**
**Concentration SMALLER THAN  3 000 µl/l (= 3 000 ppmv)**
- **Mineral oil :**
**Concentration SMALLER THAN 10 mg/kg (=10 ppmv)**
- **Total acidity : SMALLER THAN 7 µl/l (= 7 ppmv)**
**Key :  ppmv = parts per million by volume**

In the following section, we present the extraction process using of a text engineering software tool.

## 4. Extraction and processing of sentences using a text engineering approach

### 4.1 General approach
The general approach of the project is shown in the Fig. 4 below :

The first step of standard analysis and extraction is done using GATE software. It will annotate sentences containing domain concepts and extract them into an XML document.  The following steps are described in the next sub sections of the paper.

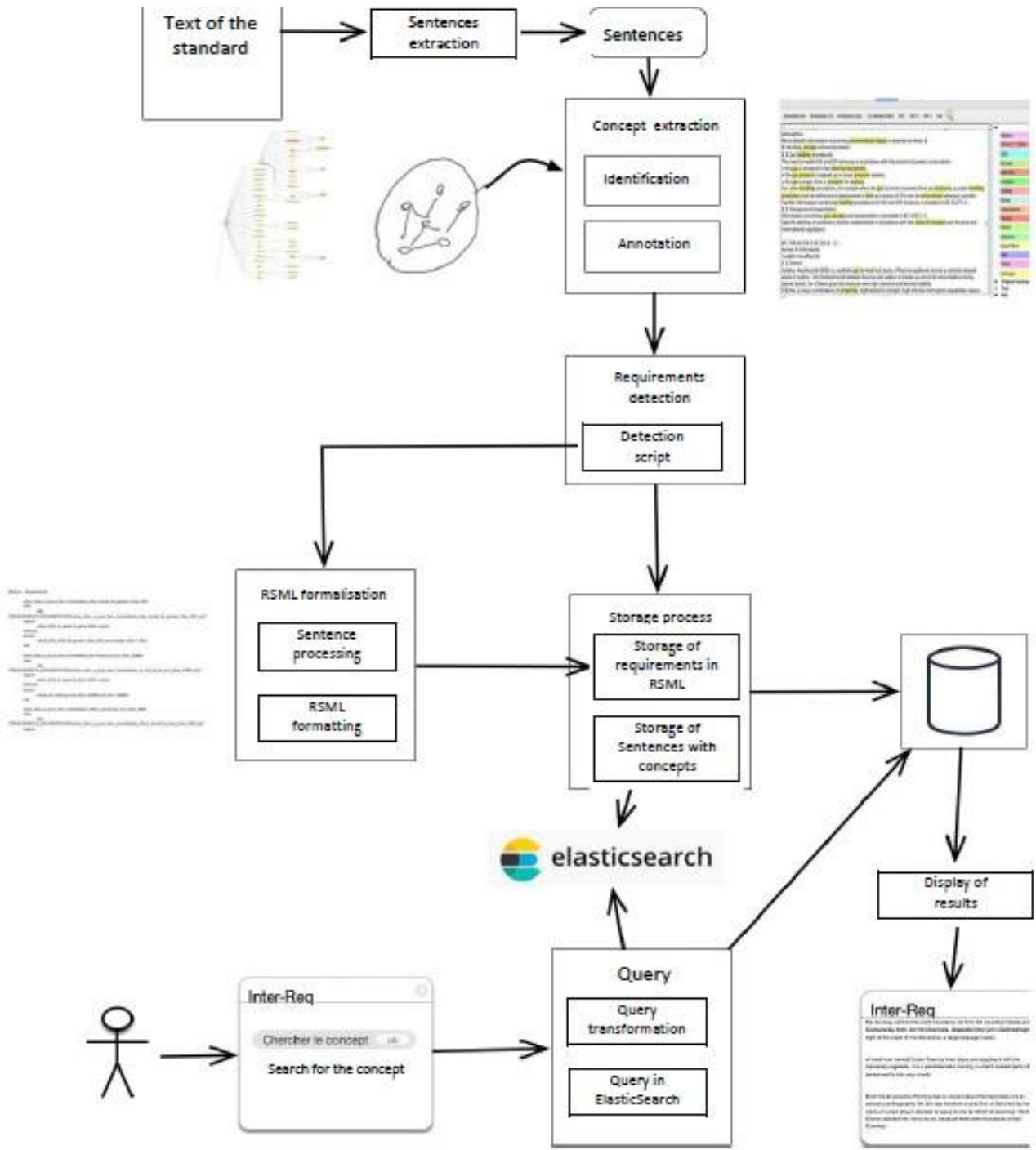

Fig. 4 : general approach of the project

**4.2 Work done with GATE**

To analyse the documents, they must first be transformed from pdf to plain text. At first, this formatting of the text will not keep the paragraphs, page layouts, etc. In a future work, we can investigate tools such as Apache's Pdf Parser or Tabula PDF to keep the layout and detect tables. For the time being, we use the analysis modules offered by GATE. The PDF thus transformed into raw text will be analysed by various modules arranged in a GATE Pipeline. To do this, we use the standard ANNIE pipeline, which allows us to tokenise the text, separate the text into sentences, have a Part-of-Speech tagger and a morphological analyser. In addition to these steps, the ontology is imported into GATE, which will annotate the text and extract the concepts corresponding to the ontology in the text.

A few JAPE rules are needed to annotate the concepts and extract the sentences, which will then allow the annotations to be exported to an XML file, below an example of one of the JAPE rules :

```
Phase: LookupRename
Input: Lookup
Options: control = appelt
Rule: RenameLookup
```

```
(
{Lookup.type == class}
):matchOntology
-->
{
AnnotationSet YtagAS = (AnnotationSet) bindings.get("matchOntology");
Annotation theLookup = (gate.Annotation)YtagAS.iterator().next();
FeatureMap feat = theLookup.getFeatures();
outputAS.add(YtagAS.firstNode(),YtagAS.lastNode(),"matchOntology",feat);
}
```

The result of the rules is given in the screenshot below (Fig. 5) :

Fig. 5 : Representation of a JAPE rule in GATE

In addition of using the ontology to detect the sentence, we created some simple JAPE rules to detect the sentences that are requirements. Currently, all requirement formulations that contain neither concepts nor these verbs are unfortunately excluded, and all sentences that contain a concept and one of the above verbs, but which are not a requirement, are retained.

These annotated sentences exported in the XML document are processed in an application written in Python which allows to extract the requirements and transform them into a RSML file. Some part of this are still done manually, as some of the grammar used in the sentences are too complex to be completely automatically transformed.

In the next section, we will apply RSML to formalise the requirements with the aim of automating the formulation of requirements

## 5  Towards an automation of requirements formalization

### 5.1 RSML formalizing of the requirements
RSML is used in our project to express requirements formally, in order to facilitate the interoperability of requirements expressions across different normative documents in the same or related domains.  Insofar as it allows the generation of the code corresponding to the modelled requirement, it can contribute to facilitate the comparison between requirements from different normative documents.

However, as will be seen later, it relies on a fairly strict expression of the original requirement before it can be transformed into code - and may therefore require a significant re-formulation of the requirement.

Below we show some of the requirements of the standard, once expressed in RSML. It should be noted that the "formal" formulation allows for a more refined exploitation than the "natural language" formulation of the requirement, particularly from the point of view of alignment / comparison between requirements from different standards. For each of the modelled requirements, the generated code files are provided :
- domain_knowledge.e
- requirements_documentation.e
- SysML.xmi
- the results files (code) of the requirements processing corresponding to the classes in the .rsml file

Below two cases of requirement processing based on the Ex 1 (text) and Ex. 3 (table) described in the section 3.3 :

**5.1.1 Requirement for responsibility** (text) : *It is the responsibility of the supplier to guarantee that the delivered gas or gas mixture is non-toxic, in accordance with international and local regulations.*
The RSML representation of the requirement using GEMOC Studio is provided Fig. 6 below :

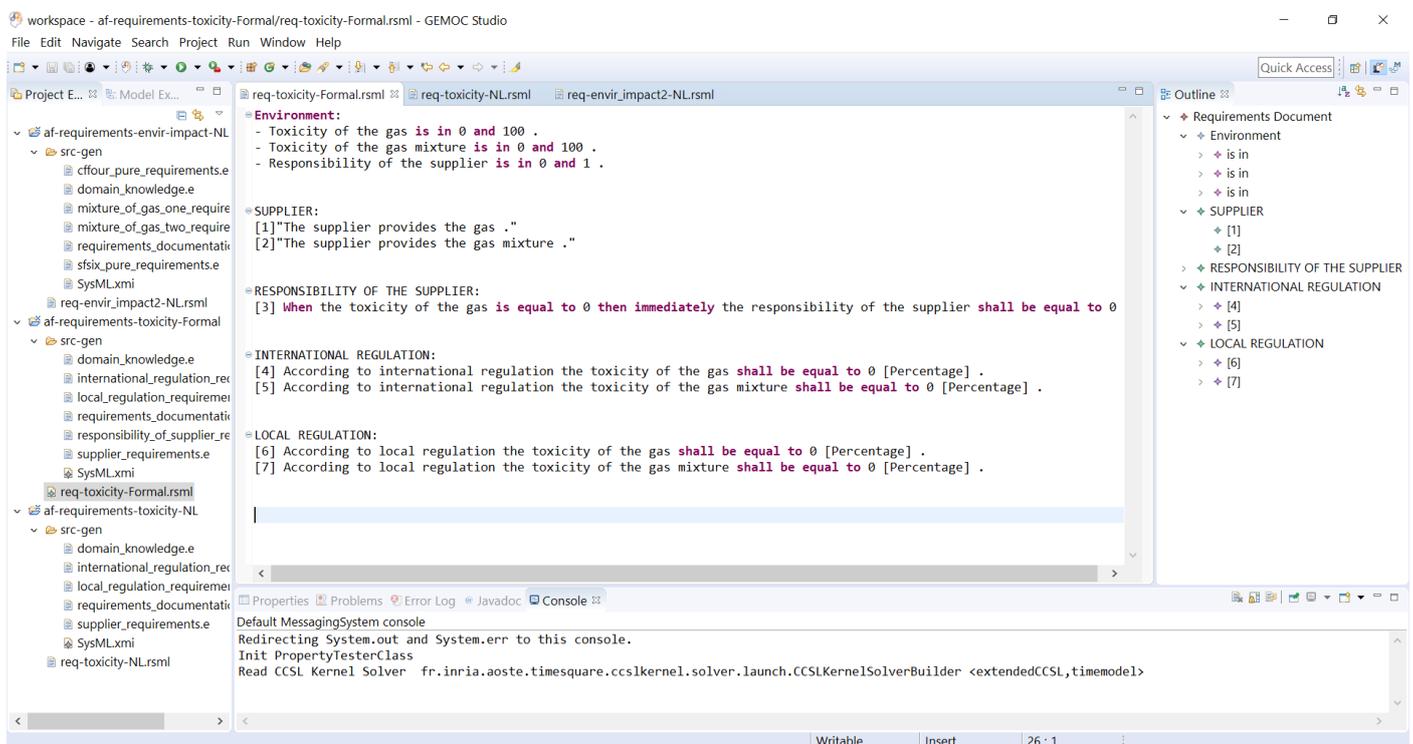

Fig. 6 : RSML representation of the requirement using GEMOC Studio

* Excerpt of the code generated (domain_knowledge.e) :

```
note
        EIS: "src=req-toxicity-Formal.rsml", "ref=Environment", "type=trace"
        description: "[
        This class contains the domain knowledge that will be used by requirements.
]"
class DOMAIN_KNOWLEDGE
feature
        toxicity_of_gas : DOUBLE
        toxicity_of_gas_mixture : DOUBLE
        responsibility_of_supplier : DOUBLE
invariant
        toxicity_of_gas_is_in_0_and_100: toxicity_of_gas > 0 and toxicity_of_gas < 100
        toxicity_of_gas_mixture_is_in_0_and_100: toxicity_of_gas_mixture > 0 and toxicity_of_gas_mixture < 100
        responsibility_of_supplier_is_in_0_and_1: responsibility_of_supplier > 0 and responsibility_of_supplier < 1
end
```

*Excerpt of the result generated : supplier_requirement.e :

```
note
        EIS: "src=req-toxicity-Formal.rsml", "ref=SUPPLIER", "type=trace"
        description: "[
        This class contains requirements in the context of: SUPPLIER.
]"
class SUPPLIER_REQUIREMENTS
inherit
        DOMAIN_KNOWLEDGE
feature
        -- For temporal requirements
        duration: DOUBLE
        -- States range
        -- States

feature -- Requirements
        requirement_1
        note
                doc: "{REQUIREMENTS_DOCUMENTATION}.requirement_1_doc"
        deferred
        end

        requirement_2
        note
                doc: "{REQUIREMENTS_DOCUMENTATION}.requirement_2_doc"
        deferred
        end
end
```

In addition of the sentences processed manually after the automatic extraction of them, we are able to semi-automatically process the tables as they are already formalized, and allows us to skip the process to transform the process of transforming the sentences.

The following example has been created with a semi-automatic script written in Python.

**5.1.2 Requirements for complementary gases to be used in SF6 mixtures** (Table 1 – section 2.2)
The RSML representation of the requirement using GEMOC Studio is provided in Fig. 7 below :

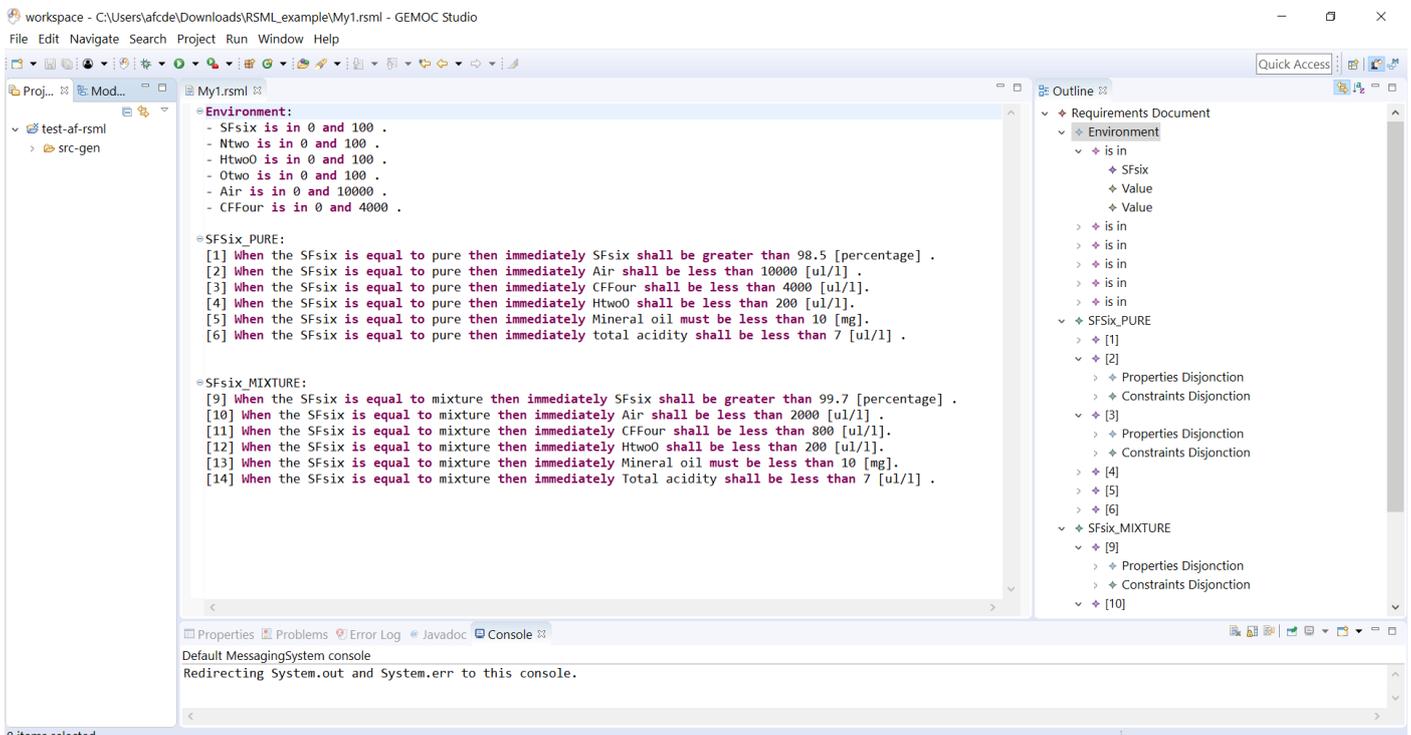

Fig. 7 : RSML representation of the requirement using GEMOC Studio

* Excerpt of the code generated (class REQUIREMENTS_DOCUMENTATION.e) :

```
note
        EIS: "src=My1.rsml", "type=trace"
        description: "[
                This class contains the documentation of all requirements.
        ]"
class REQUIREMENTS_DOCUMENTATION
feature -- Requirements
        when_sfsix_is_pure_then_immediately_sfsix_should_be_greater_than_98.5_doc: STRING
                note
                        EIS: "src=My1.rsml", "ref=[1]", "type=trace"
                        doc: "true"
                do
                        Result := "[
                                [1] When the SFsix is equal to pure then immediately SFsix shall be greater than 98.5 [percentage] .
                                ]"
                end

        when_sfsix_is_pure_then_immediately_air_should_be_less_than_10000_doc: STRING
                note
                        EIS: "src=My1.rsml", "ref=[2]", "type=trace"
                        doc: "true"
                do
                        Result := "[
                                [2] When the SFsix is equal to pure then immediately Air shall be less than 10000 [ul/l] .
                                ]"
                end
```

*Excerpt of the result generated : SF6 mixture requirement :

```
note
        EIS: "src=My1.rsml", "ref=SFsix_MIXTURE", "type=trace"
        description: "[
        This class contains requirements in the context of: SFsix_MIXTURE.
]"
class SFSIX_MIXTURE_REQUIREMENTS
inherit
        DOMAIN_KNOWLEDGE
feature
        -- For temporal requirements
        duration: DOUBLE

        -- States range
        mixture : DOUBLE = 1

        -- States
        cffour : DOUBLE
        mineral_oil : DOUBLE
        sfsix : DOUBLE
        air : DOUBLE
        htwoo : DOUBLE
        total_acidity : DOUBLE

feature -- Requirements

        when_sfsix_is_mixture_then_immediately_sfsix_should_be_greater_than_99.7
        note
                doc:
"{REQUIREMENTS_DOCUMENTATION}.when_sfsix_is_mixture_then_immediately_sfsix_should_be_greater_than_99.7_doc"
        require
                when_sfsix_is_equal_to_mixture: (sfsix = mixture)
        deferred
        ensure
                check_sfsix_shall_be_greater_than_99.7_percentage: (sfsix > 99.7)
        end
```

In order to be able to query these sentences and to accurately retrieve requirements containing a concept, in this last step, an application in the form of a Python service was implemented. This application consists of two separate modules. The module for indexing sentences uses an Elasticsearch server: Elasticsearch[4] [23] is a system using Lucene to index and search data. Like Lucene and SOLr, it is able to perform faceted queries, which is very useful in the case of this application, as we have several possible facets.

One of the facets we want to exploit here is the concept defined in the sentences corresponding to requirements. We also define another facet which is the type of sentence, in the case of this application, this facet will be by default the requirement. For indexing, Elasticsearch allows to manage and implement its own data structures, corresponding to the requirements. In the case of this application, the data is indexed according to concepts and sentences. The concept is indexed separately from the sentence.

**5.2 Storage and indexing processes**

To index documents in elasticsearch, one must first create an elasticsearch server and connect to it. In a second step, several methods are implemented to index the data. Only sentences that contain at least one concept and one of the verbs will be indexed. In order to allow for more flexibility and to integrate other types of data in the future, the type of sentences is not permanent, but it can be specified in the method.

For the moment each sentence is indexed one by one, in the future a reverse index can be considered, if the number of sentences becomes too large.

The module that allows the data to be queried is implemented as a service. Several ways of querying the data have been implemented: the first one is the one that allows to have all the documents. The second allows you to create your query with specific parameters, which are the concepts. Three APIs have been implemented, one allowing to retrieve all the documents indexed in the Elasticsearch server, one allowing to create a query containing a concept, and another allowing to query a document according to its identity. In the future, it will be possible to implement as many access points as necessary. These three services were implemented using Flask, a Python framework.

\* Excerpt of the ElasticSearch code :

```
class ElasticSearchS:
    def __init__(self, index_name):
        self.index_name = index_name
    # connect to elastisearch
    def connect_elasticsearch(self):
        es = Elasticsearch([{'host': 'localhost', 'port': 9200}])
        if es.ping():
            print('It is connected')
        else:
            print('Seems to be not connected')
        return es

    def search_all_document(self):
        es = self.connect_elasticsearch()
        resultat = es.search(index="document_norms", body={'size': 10000, 'query': {'match_all': {}}})
        return resultat
```

**5.3 Queries – viewing results**

A graphical interface can be implemented on top of these services, which will allow the future user to create queries in natural language. The user will also be able to have different types of visualisation of the results of his queries according to his needs (Kibana). Due to lack of time, the queries have not been fully developed. The queries are basic, they are developed in service, and the user interface is missing, see Fig. 8 below an example of query.

---

[4] https://www.elastic.co/fr/what-is/elasticsearch

```
                    }
                },
                {
                    "_index": "document_norms",
                    "_type": "_doc",
                    "_id": "_4AcqnYBM2AVOHZE6JY6",
                    "_score": 1.8460661,
                    "_source": {
                        "type": "exigence",
                        "title": "IEC 60376 ED3",
                        "sentence": "In that case, to limit the total uncertainty after a typical gas handling operation such as re-filling in order to comply with OEM specifications on
the mixture composition ratio, higher SF6 purity grade > 99,7 % volume shall be used.",
                        "concept": "SF6",
                        "date": "2020-12-28T17:09:49.882388"
                    }
                },
                {
                    "_index": "document_norms",
                    "_type": "_doc",
                    "_id": "A4AcqnYBM2AVOHZE6ZcK",
                    "_score": 1.8460661,
                    "_source": {
                        "type": "exigence",
                        "title": "IEC 60376 ED3",
                        "sentence": "Due to this impact, SF6, CF4 and their mixture gas shall be handled carefully to prevent deliberate release of SF6 and CF4 gas into the atmosphere.",
                        "concept": "SF6",
                        "date": "2020-12-28T17:09:50.090412"
                    }
                },
                {
                    "_index": "document_norms",
                    "_type": "_doc",
                    "_id": "CIAcqnYBM2AVOHZE6pdt",
                    "_score": 1.8460661,
                    "_source": {
                        "type": "exigence",
                        "title": "IEC 60376 ED3",
                        "sentence": "Due to this impact, SF6, CF4 and their mixture gas shall be handled carefully to prevent deliberate release of SF6 and CF4 gas into the atmosphere.",
                        "concept": "SF6",
                        "date": "2020-12-28T17:09:50.444584"
                    }
                },
                {
                    "_index": "document_norms",
                    "_type": "_doc",
                    "_id": "EYAcqnYBM2AVOHZE7Jec",
                    "_score": 1.8460661,
                    "_source": {
                        "type": "exigence",
                        "title": "IEC 60376 ED3",
                        "sentence": "For other handling procedures, for example when the gas has to be recovered from an enclosure, a proper handling procedure shall be defined and
implemented to limit any release of SF6 into the environment wherever possible.",
                        "concept": "SF6",
```

Fig. 8 : example of query

**Conclusion – issues - perspectives**

At the end of this paper, we realise that writing a normative requirement is not, contrary to what one might expect, and contrary to the commonly accepted opinion about writing a normative document, a neutral operation : the vocabulary used in the formulation of the requirement is not neutral, insofar as it proposes a formulation and probably also anticipates the response / treatment that will be made of the requirement sentence.
For example: if we consider the requirement on the environmental impact :
*"SF6, CF4 and SF6 mixtures with N2 and/or CF4 have a **certain** environmental impact. Due to this impact, SF6, CF4 and their mixture gas shall be handled **carefully** to prevent **deliberate** release of SF6 and CF4 gas into the atmosphere."*
- the use of bolded words presupposes that the person reading the requirement will "understand" what, for example, "carefully" means !
- but it is probably another person who will "understand" -- and adapt -- the term "certain".
- And so, what does the term "deliberate" mean in this case ?
　　　　* Should we then anticipate the possibility of a *deliberate* malicious action ?
　　　　* And if so, should it be dealt with (anticipated ?) and how should it be "dealt with" ?
Further work will be necessary to adapt the writing of the requirements to the ISO/IEC Directives (and vice versa !) in order to avoid any kind of mis-interpretation and ambiguïty. if not, the requirements could be declared as non-compliant with the Directives and thus, not valid.

From the point of view of the processing of requirements by NLP approaches, we realise how much information is implied, implicit, incomplete, but also specific to one (or more) professions likely to apply it, such as purchasing, processing, packaging, delivery of SF6 gas, or CF4, or other !
On the other hand, in this paper, we focused the study on IEC 60376, without linking it to related standards that are referenced (or not) in IEC 60376: the purpose of the project was not to establish relationships between standards that share common concepts, but to focus on the way of extracting requirements definition information from a single document, using automatic language processing techniques. Similar work on related normative documents, and then alignment of the requirements mentioned in the different normative documents, would help validate these requirements.

The International Electrotechnical Commission (IEC) Strategic Group SG 12 has published a model for various degrees of maturity and utility of « digital standards » (Figure 9).

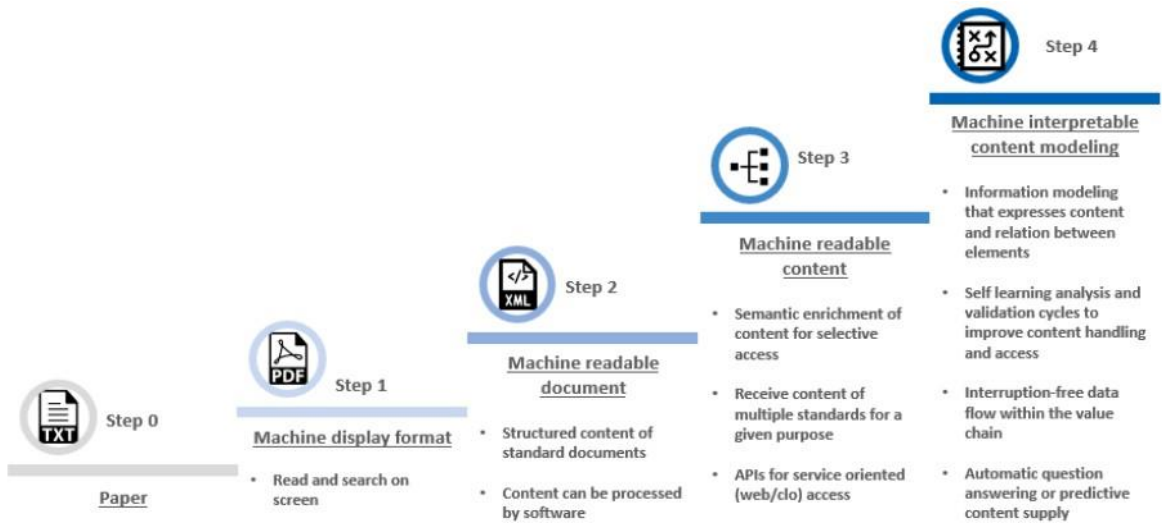

Figure 9 : IEC SG 12 Model for digital maturity and utility of standards (IEC SG 12)

Standards are currently published as documents on paper and pdf formats (Step 0 and 1). They are primarily document-centric in use, although the content actually contains large numbers of specific, granular content like definitions, essential information, governing statements (requirements, recommendations and permissions), figures and equations etc. Some standards are published by ISO and CEN using the NISO STS XML format (Step 2) [24], but the markup is limited to marking up entire clauses as single items. These single items contain multiple statements (requirements, recommendations, permissions and associated information) which is quite impractical for the end-user to relate to in a structured process (the need is one by one statement).

There is also a growing focus on publication of content in standards at Step 4 : data-centric, semantic format. Currently, the major standards development organizations (SDOs) have not yet defined formats suited for this task and then of course not developed any authoring, administration and marketing tool for such products. There is without doubt a significant innovation potential herein. However, it can be noted that, as of today, several standardisation bodies at the European or international level are beginning to work on the concept of online collaborative authoring.

Regardless of Step 3 and 4, there is a fundamental lack of a single, effective global coordination and "ownership" of terms (including symbols) and their definitions. Additionally, the commonly used authoring tool (Microsoft Word) allows full (too much !) freedom for authors to enter and structure content the way they want – not necessarily according to the rules for standards authors.

The results are :
- Terms are not commonly understood and used (by neither humans nor machines) ;
- Statements are made in an ambiguous way that cannot be handled and implemented in an efficient and correct way by the end-user (neither human nor machine).

More generally, an application of extraction of technical information from normative documents, based on information coming from different technical sources (other standards, and/or other parts of the same standard) has been carried out on a standard developed by the ISO TC 184 SC4-SC5 JWG8 committee, in the domain of production management information [25]. This work has already highlighted a number of inconsistencies and even errors in the drafting of normative documents.

The requirement, in its formulation, its content (explicit or implicit) is at the crossroad of the different professions that are likely to apply it [26], [27]. Its automatic processing presupposes the integration of the vocabularies, approaches and working methods of the professionals - without this synthesis of the knowledge it represents, this automatic extraction can be difficult to implement - and then to apply !

This constitutes a fundamental step towards the creation of future machine readable standards [28].

In terms of machine learning : at our knowledge, we didn't find any annotated dataset that could be used to extract the requirements from text. To use a precise and accurate extracting system, it is more and more proven that a generalised dataset is not so accurate [18] for very specific dataset, it is one of the reason behind the use of more standard NLP approaches.

**Acknowledgements** : Réseau de Transport d'Electricité (RTE)

**Declaration of competing interest** : the authors declare that they have no known competing financial interests or personal relationships that could have appeared to influence the work reported in this paper.